\documentclass[11pt]{article}
\usepackage{amssymb,epsf,graphics,graphicx}
\newcommand{\ds}{\displaystyle}

\newcommand\blank[1]{}

\newcommand{\fract}[2]{{\textstyle\frac{#1}{#2}}}

\newcommand{\ri}{\right}
\newcommand{\ep}{\varepsilon}
\newcommand{\lf}{\left}
\newcommand{\te}{\theta}

\newcommand\Rth{{\mathbb R}}

\newcommand\eq{\begin{equation}}
\newcommand\en{\end{equation}}
\newcommand\bea{\begin{eqnarray}}
\newcommand\eea{\end{eqnarray}}
\newcommand\nn{\nonumber}
\newcommand\ba{\(\begin{array}}
\newcommand\ea{\end{array}\)}

\newcommand{\vev}[1]{\langle\,#1\,\rangle}
\newcommand{\resection}[1]{\setcounter{equation}{0}\section{#1}}

\newcommand{\ket}[1]{\ensuremath{\mbox{\normalsize $| #1\rangle$}}}

\newcommand{\One}{{\hbox{{\rm 1{\hbox to 1.5pt{\hss\rm1}}}}}}
\newcommand{\cH}{{\cal H}}
\newcommand{\IJMP}[1]{Int.\ J.\ Mod.\ Phys.\ {\bf #1}}

\newcommand{\NP}[1]{Nucl.\ Phys.\ {\bf #1}}
\newcommand{\PL}[1]{Phys.\ Lett.\ {\bf #1}}

\newcommand{\PRL}[1]{Phys.\ Rev.\ Lett.\ {\bf #1}}

\begin{document}
\begin{titlepage}
\vskip 0.5cm
\begin{flushright}
DCPT-04/11 \\
SPhT-T04/037 \\
{\tt hep-th/0404014} \\
\end{flushright}
\vskip .5cm
\begin{center}
{\Large{\bf Integrable quantum field theory with boundaries: \\
the exact g-function }}
\end{center}
\vskip 0.8cm
\centerline{Patrick Dorey$^{1,2}$, Davide Fioravanti$^3$,
Chaiho Rim$^4$ and Roberto Tateo$^{1,5}$}
\vskip 0.9cm
\centerline{${}^1$\sl\small Dept.~of Mathematical Sciences,
University of Durham,}
\centerline{\sl\small  Durham DH1 3LE, United Kingdom\,}
\vskip 0.3cm
\centerline{${}^2$\sl\small Service de Physique Th{\'e}orique, CEA-Saclay,}
\centerline{\sl\small F-91191 Gif-sur-Yvette Cedex, France}
\vskip 0.3cm
\centerline{${}^3$\sl\small  Dept.~of Mathematics, University of
York,}
\centerline{\sl\small  York YO10 5DD, United Kingdom}
\vskip 0.3cm
\centerline{${}^4$\sl\small Dept. of Physics, Chonbuk National
University,}
\centerline{\sl\small Chonju 561-756, Korea}
\vskip 0.2cm
\centerline{${}^{5}$\sl\small Dip. di Fisica Teorica and INFN,
Universit\`a di Torino,}
\centerline{\sl\small Via P. Giuria 1, 10125 Torino, Italy}
\vskip 0.2cm
\centerline{E-mails:}
\centerline{ p.e.dorey@durham.ac.uk, df14@york.ac.uk,
rim@chonbuk.ac.kr,  tateo@to.infn.it}

\vskip 1.25cm
\begin{abstract}
\noindent
The $g$-function was introduced by Affleck and Ludwig in the
context of critical quantum systems with boundaries. In the
framework of the thermodynamic Bethe ansatz (TBA) method for
relativistic scattering theories, all attempts to write an exact
integral equation for the off-critical version of this quantity have,
up to now, been unsuccesful. We tackle
this problem by using an  n-particle cluster
expansion,
close in spirit to form-factor calculations of
correlators  on the plane. The leading contribution
already disagrees with all previous
proposals, but a study of this and subsequent terms allows us
to deduce an exact infrared
expansion for $g$, written purely in  terms of
TBA pseudoenergies. Although we only treat the
thermally-perturbed Ising and the scaling Lee-Yang models in
detail, we  propose a general formula for $g$ which should be
valid for any model with entirely diagonal scattering.
\end{abstract}
\vskip 0.2cm
\leftline{\small Keywords: Boundary problems; Conformal field theory; Integrability; Thermodynamic Bethe ansatz }
\leftline{\small PACS: 05.20.-y; 11.25.Hf; 11.55.Ds; 68.35.Rh}

\end{titlepage}
\setcounter{footnote}{0}
%%%%%%%%%%%%%%%%%%%%%%%%%%%%%%%%%%%%%
\def\thefootnote{\fnsymbol{footnote}}
%%%%%%%%%%%%%%%%%%%%%%%%%%%%%%%%%%%%%%%%%%%%%%%%%%%%%%%%%%%%%%%%%%%%%
%%%  start of the paper  %%%%%%%%%%%%%%%%%%%%%%%%%%%%%%%%%%%%%%%%%%%%
%%%%%%%%%%%%%%%%%%%%%%%%%%%%%%%%%%%%%%%%%%%%%%%%%%%%%%%%%%%%%%%%%%%%%
%
\resection{Introduction}
\label{intr}
The study of two-dimensional conformal field theories with
boundaries~\cite{cardy} and their integrable
perturbations~\cite{Zam1,GZ,BLZ1} is of
interest both in condensed matter physics~\cite{Sa1} and in
string theory~\cite{AS}. An
important quantity emerging from the definition of
the cylinder partition function for these field theories is
the $g$-function, the
`ground-state degeneracy' or `boundary entropy', which for models
critical in the bulk was introduced some
years ago by Affleck and Ludwig~\cite{AL}. While many
interesting questions remain in these cases~\cite{FK,Kon}, in this
paper we shall deal with the further issues which arise for
off-critical, massive,  systems.

The $g$-function for massive field theories can be defined
as follows~\cite{LMSS,Ch,Us1,Us3}. There are two possible
Hamiltonian descriptions of the cylinder partition function. In the
so-called L-channel representation the r{\^o}le of time
is taken by $L$, the circumference of the circle:
\eq
  Z_{\alpha\beta}
= {\rm Tr}_{\cH_{(\alpha,\beta)}} e^{-LH_{\alpha\beta}^{\rm strip}(M,R)} =
\ds{  \sum_{n=0}^{\infty}\, \,
 e^{ - L E_n^{\rm strip}(M,R)}
}\;.
\label{rrchan}
\en
In this formula,
$H_{\alpha\beta}^{\rm strip}$
propagates states in $\cH_{(\alpha,\beta)}$,
the Hilbert space for an
interval of length $R$ with boundary conditions $\alpha$ and $\beta$
imposed at the two ends,
$E_n^{\rm strip} \in\,{\rm spec}(H_{\alpha\beta}^{\rm strip})$\,, and
$M$ is the mass of the lightest  particle in the
theory. In the R-channel representation the r{\^o}le of time is
instead taken by $R$, the length of the cylinder:
\eq
  Z_{\alpha\beta}
=   \vev{\alpha|
  \,e^{-RH^{\rm circ}(M,L)}\,
  |\beta}  =\sum_{n=0}^{\infty}
 ~~
{\cal G}_{\alpha}^{(n)}(l) {\cal G}_{\beta}^{(n)}(l)
\, e^{ - R E_n^{\rm circ}(M,L) },~~(l=ML)
\label{llchan}
\en
where  $E_n^{\rm circ} \in \,{\rm spec}(H^{\rm circ})$ and
\eq
{\cal G}_{\alpha}^{(n)}(l)=
\fract{\vev{\alpha|\psi_n}}{\vev{\psi_n|\psi_n}^{1/2} }~.
\en
In equation (\ref{llchan}),
the boundary states $\ket{\alpha}$, $\ket{\beta}$
and the  eigenbasis $\{\ket{\psi_n}
\}$ of the Hamiltonian $H^{\rm circ}$ have been used. These
are defined on a circle  of  circumference $L$ and
propagate along the `time' direction $R$. At large $l$, the
function $\ln {\cal G}_{\alpha}^{(0)}(l)$
 grows linearly:
\eq
\ln {\cal G}_{\alpha}^{(0)}(l) \sim - f_{\alpha} L~,
\en
where the constant $f_{\alpha}$ contributes to the constant (boundary)
part of the ground-state energy on the strip (see eq.~(\ref{e0strip})).
The standard $g$-function  is then defined as
\eq
\ln g_{\alpha}(l)= \ln {\cal G}_{\alpha}^{(0)}(l) + f_{\alpha} L~.
\label{gfdef}
\en
In theories with only massive excitations in the bulk,
$\ln g_{\alpha}(l)$ tends exponentially  to zero at large $l$.

\vskip 8pt
\[
\begin{array}{c}
\refstepcounter{figure}
\label{fig:lchan}
\epsfxsize=.60\linewidth
\epsfbox{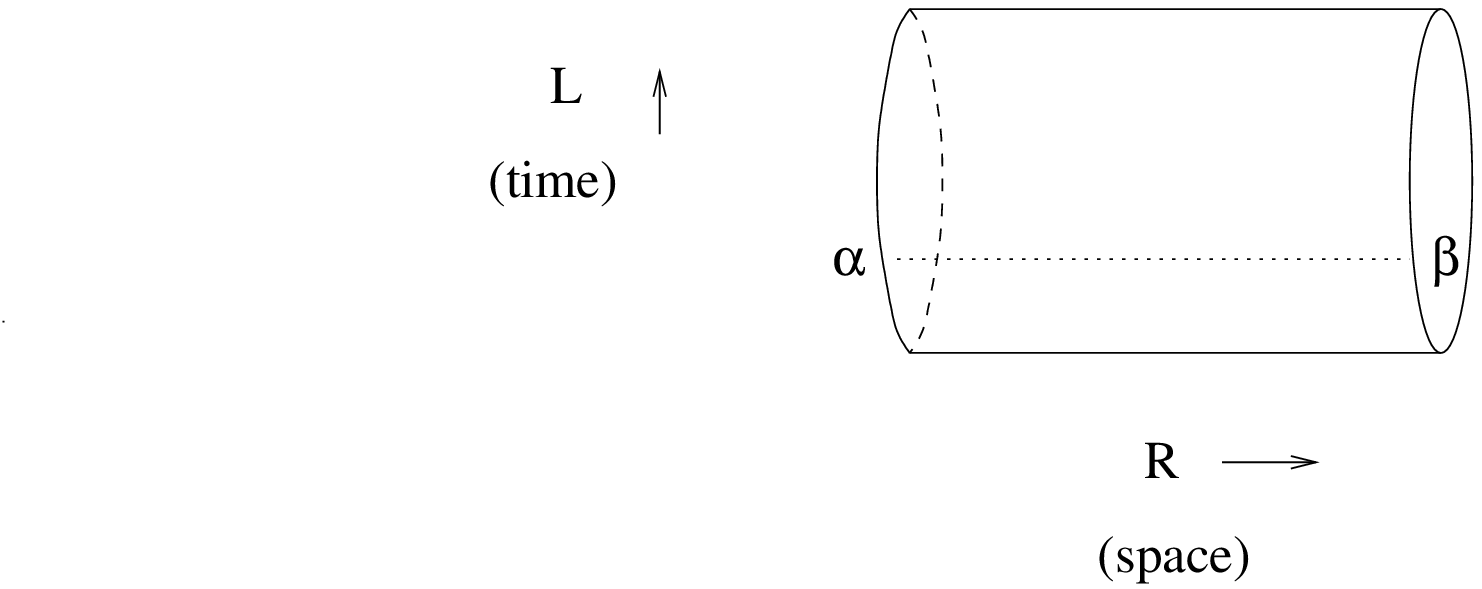}\qquad
\\
\parbox{.9\linewidth}{\small \raggedright
Figure \ref{fig:lchan}:~The L-channel decomposition; states
$\ket{\chi_n}$ live on the dotted line segment along the
cylinder. }
\end{array}
\]

\[
\begin{array}{c}
\refstepcounter{figure}
\label{fig:rchan}
\epsfxsize=.60\linewidth
\epsfbox{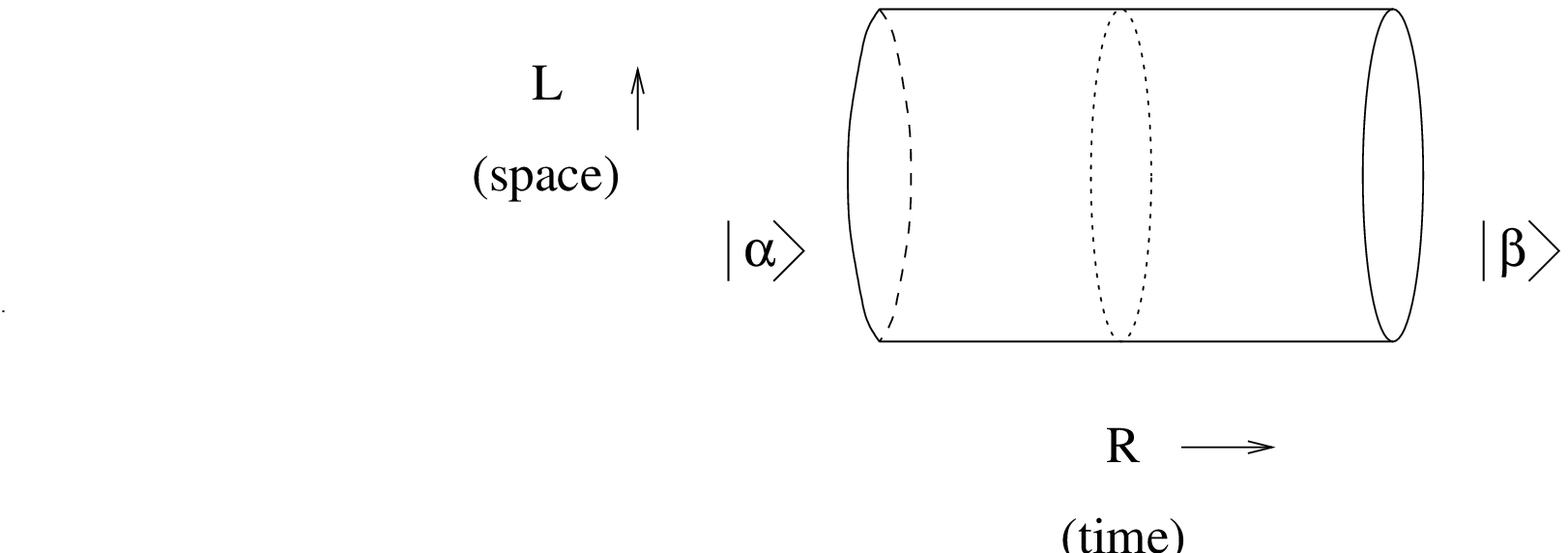}\qquad
\\
\parbox{.9\linewidth}{\small \raggedright
Figure \ref{fig:rchan}:~The R-channel decomposition;
states $\ket{\psi_n}$ live on the dotted circle around the cylinder. }
\end{array}
\]

\noindent
The two decompositions are illustrated in figures \ref{fig:lchan} and
\ref{fig:rchan}.

The equality of  (\ref{rrchan}) and
(\ref{llchan}) results in the following important identity:
\eq
\sum_{n=0}^{\infty}\, \,
 e^{ - L E_n^{\rm strip}(M,R)}
=
\sum_{n=0}^{\infty}
~~ {\cal G}_{\alpha}^{(n)}(l) {\cal G}_{\beta}^{(n)}(l)
\, e^{ - R E_n^{{\rm circ}}(M,L) }.
\label{Pid}
\en
The purpose of this paper is to develop
an exact expression for the ground-state function
$\ln {\cal G}^{(0)}_{\alpha}(l)$ through the large-$R$ limit of
(\ref{Pid}), with boundary conditions $\beta=\alpha$.
As it stands, the fact that $E_0^{\rm circ}(M,L)$
is negative makes the RHS
of (\ref{Pid}) diverge as $R\to\infty$; however, rearranging gives
\bea
2 \ln  {\cal G}_{\alpha}^{(0)}(l) &=&
R   E_0^{\rm circ}(M,L) - L  E_0^{\rm strip}(M,R)
\label{gexp1} \\
&+&
\ln \left ( 1+
\sum_{n=1}^{\infty}\, \,
 e^{ - L (E_n^{\rm strip}(M,R)-E_0^{\rm strip}(M,R))} \right )
+O(e^{-R (E_1^{\rm circ}-E_0^{\rm circ})})~. \nn
\eea
We shall restrict our attention to massive theories with
non-degenerate ground state on the plane. For
these models the non-zero mass gap gives the final term
the leading behaviour
\eq
O(e^{-R (E_1^{\rm circ}(M,L)-E_0^{\rm circ}(M,L))})
\sim O(e^{-RM})
\label{oo1}
\en
in the domain   $R\gg L\gg 0$.
In this same domain, $E^{\rm strip}_0(M,R)$ tends to its limiting
form as
\eq
E_0^{\rm strip}(M,R)=
{\cal E}M^2R+2f_{\alpha}+O(e^{-RM})
\en
where ${\cal E}$ and $f_{\alpha}$ are the extensive
bulk and boundary free
energies, as in (\ref{e0strip}).
These constraints are crucial for the validity of the perturbative
treatment to be introduced  in the following sections:
the higher corrections have a clear
dependence on $R$
and  do not contribute to
the $g$-function.
Discarding
these exponentially-suppressed terms and using the definition
(\ref{gfdef}), we finally obtain
\eq
2 \ln  g_{\alpha}^{(0)}(l) \sim
R  \Bigl( E_0^{\rm circ}(M,L) - {\cal E}M^2L\Bigr)
+\ln\!\left( 1+
\sum_{n=1}^{\infty}\, \,
 e^{ - L (E_n^{\rm strip}(M,R)-E_0^{\rm strip}(M,R))} \right).
\label{gexp2}
\en
Having taken $R$ to be large, the cluster expansion involves letting $L$
tend to infinity as well, so that an expansion of the RHS of
(\ref{gexp2}) can be developed in terms of one-, two- and so on particle
contributions, which themselves can be consistently
estimated using the Bethe-ansatz approximated levels
(\ref{e2}), (\ref{BAEQ1}).
Note that this differs from the strategy adopted
in \cite{LMSS}, where a saddle-point evaluation of the dominant
contributions at finite $L$ was made instead.

The rest of this paper is organised as follows. In section~\ref{ising}
the cluster method is exemplified by studying the free fermion
theory associated to the thermally-perturbed Ising model. The
resulting integral expression for $\ln g_{\alpha}(l)$
turns out to be in full agreement with
previous results of \cite{LMSS,Ch}.  In
section~\ref{earlier} two previous proposals~\cite{LMSS,RC}
for $\ln g_{\alpha}(l)$ are described  and in
section~\ref{ly11} the scattering data for the scaling Lee-Yang model,
our working interactive example, are summarised. The
ultraviolet result obtained from the conformal perturbation theory and
the boundary truncated conformal space approximation
(BTCSA)~\cite{Us1,Us3,Kon} is compared with
infrared numerical results from the
Bethe Ansatz, and the equivalence between the
two functions is confirmed by a large overlap at intermediate scales.

This agreement motivates the search for an exact analytic
expression. This is  the main objective of sections~\ref{onep},
\ref{twop} and
\ref{exact} where the large strip-width
($R \rightarrow \infty$) limit
is explicitly taken and sums over the quantum numbers are transformed
into integrals in rapidity variables. This analysis
leads to the final exact expansion for $\ln g_{\alpha}(l)$ given
in equation (\ref{finalc}). This and its generalization
(\ref{finald}) constitute the main results of the paper. In section
\ref{exact} we also briefly comment on the similarity
between our results and one recently obtained by Woynarovich in
\cite{Woy}. Section \ref{conclusions} contains  our
conclusions. Finally, in  Appendix~\ref{TBABA}
we summarise the main equations used to develop our programme: the
thermodynamic Bethe ansatz~\cite{AlZam1} and the
Bethe quantisation conditions. In
Appendix~\ref{apps} the reflection factors for the Ising model are
recalled  and  explicit expressions  for the
 boundary entropy  for  free-free and   fixed-fixed
 conditions   are presented.

\resection{A simple example: the Ising model}
\label{ising}
We start with the study of the free Majorana fermion theory
corresponding to the thermally-perturbed Ising model on a
strip. The partition function is (cf. \cite{Ch})~\footnote{Notice
that the zero momentum ($\te_j=0 \leftrightarrow j=0$)
particle state is forbidden.} :
\eq
Z_{\alpha  \alpha}= e^{-L E_0^{\rm strip}(M,R)}
\prod_{j>0} \left(1+e^{-l \cosh \te_j} \right)~,~~~(l=ML)~,
\label{zz}
\en
or
\bea
\ln Z_{\alpha \alpha}&=&-L E_0^{\rm strip}(M,R) +
\sum_{j>0} \ln(1+e^{-l \cosh \te_j}) \cr
&=& -L E_0^{\rm strip}(M,R) + {1 \over 2} \sum_{j=-\infty}^{\infty}
\ln(1+e^{-l \cosh
\te_j})- {1 \over 2} \ln( 1+e^{-l})\,.
\label{logzz1}
\eea
Due to the singular behaviour for the Ising model
of the bulk and linear  terms
${\cal E}$ and $f_\alpha$ defined in (\ref{ecircl}) and
(\ref{e0strip}) below, it is convenient, exceptionally for this case,
to work with subtracted energies tending exponentially to
zero at large scales:
\eq
E_0^{\rm strip}(M,R) \Big|_{R \gg 1}  \sim 0
\label{eostripi}
\en
and consistently to set
\eq
   E_0^{\rm circ}(M,L) = -
   \int_{\Rth} {d \te \over 2\pi} \,
   M \cosh\theta\, \ln \big(1+e^{-l \cosh \theta} \big)\,
\label{ecirising}
\en
(cf.~(\ref{ecircl})).
 Starting from the quantization condition
\eq
r \sinh \te_ j -i \ln R_{\alpha}(\te_j) =
\pi j~,~~~(r=MR)
\label{BAi}
\en
with integer $j$  and $R_{\alpha}(\te_j)$ as
defined in (\ref{Rkk}), writing equation (\ref{BAi}) with $j
\rightarrow j+1$ and subtracting (\ref{BAi})
from the result, we find in the large $R$ limit
\eq
{\Delta \te_j \over \pi} \left( r  \cosh(\te_j) -i
 {d
\over d
\te}\ln R_{\alpha}(\te_j)\right)+
O((\Delta\theta_j)^2)=1\,.
\en
Substituting  this into (\ref{logzz1}),
\eq
\ln Z_{\alpha \alpha} \sim  {1 \over 2  } \int_{\Rth}
d\te \left({r \over \pi}  \cosh(\te) +
\phi_{\alpha}(\te)  -  \delta(\te) \right) \ln (1+e^{-l \cosh
\te})\,,
\label{logz0}
\en
where $\phi_{\alpha}(\te)$ is  given in (\ref{phikk}). In
the latter equation we recognize a part corresponding to $R
E_0^{\rm circ}(M,L)$, and, considering also (\ref{eostripi}),
we arrive at the exact result
\eq
2 \ln g_{\alpha}(l) = \lim_{R \rightarrow \infty}
(\ln Z_{\alpha \alpha}(L,R)+ R E_0^{\rm circ}(M,L))= { 1 \over 2}
\int_{\Rth} d
\te
 \left( \phi_{\alpha}(\te)-  \delta(\te) \right)
 \ln(1+ e^{-l \cosh \te})\,.~~~~~~
\label{iex}
\en
This  coincides with the result found in~\cite{LMSS,Ch}
using a different technique. The match confirms
the correctness of our method, at least in this case,
and motivates its study in more complicated models.

In figures \ref{ifix} and \ref{ifree} the integration  of
(\ref{iex}) for free-free and fixed-fixed boundary
conditions,
corresponding to $k=1$ and $k=-\infty$ in  (\ref{Rkk}) and
(\ref{phikk})\footnote{
By studying the monodromies of the integral (\ref{iex}), we have also
found  more explicit expressions for $\ln g_{\bf
fixed}(l)$ and $\ln g_{\bf free}(l)$; these are
given in Appendix~\ref{apps}.},
is compared  with  numerical results obtained by
estimating the large-$R$  partition function
(\ref{zz}) using  the   Bethe ansatz quantized energy
levels~(\ref{BAi}) directly, and then extracting the boundary entropy
through the relation
\eq
2 \ln g_{\alpha}(l) \sim
\big(\ln Z_{\alpha \alpha}(L,R)+
R E_0^{\rm circ}(M,L) \big) \Big|_{r\gg 1}~.
\en
\[
\begin{array}{cc}
\refstepcounter{figure}
\label{ifix}
\epsfxsize=.49\linewidth
\epsfbox{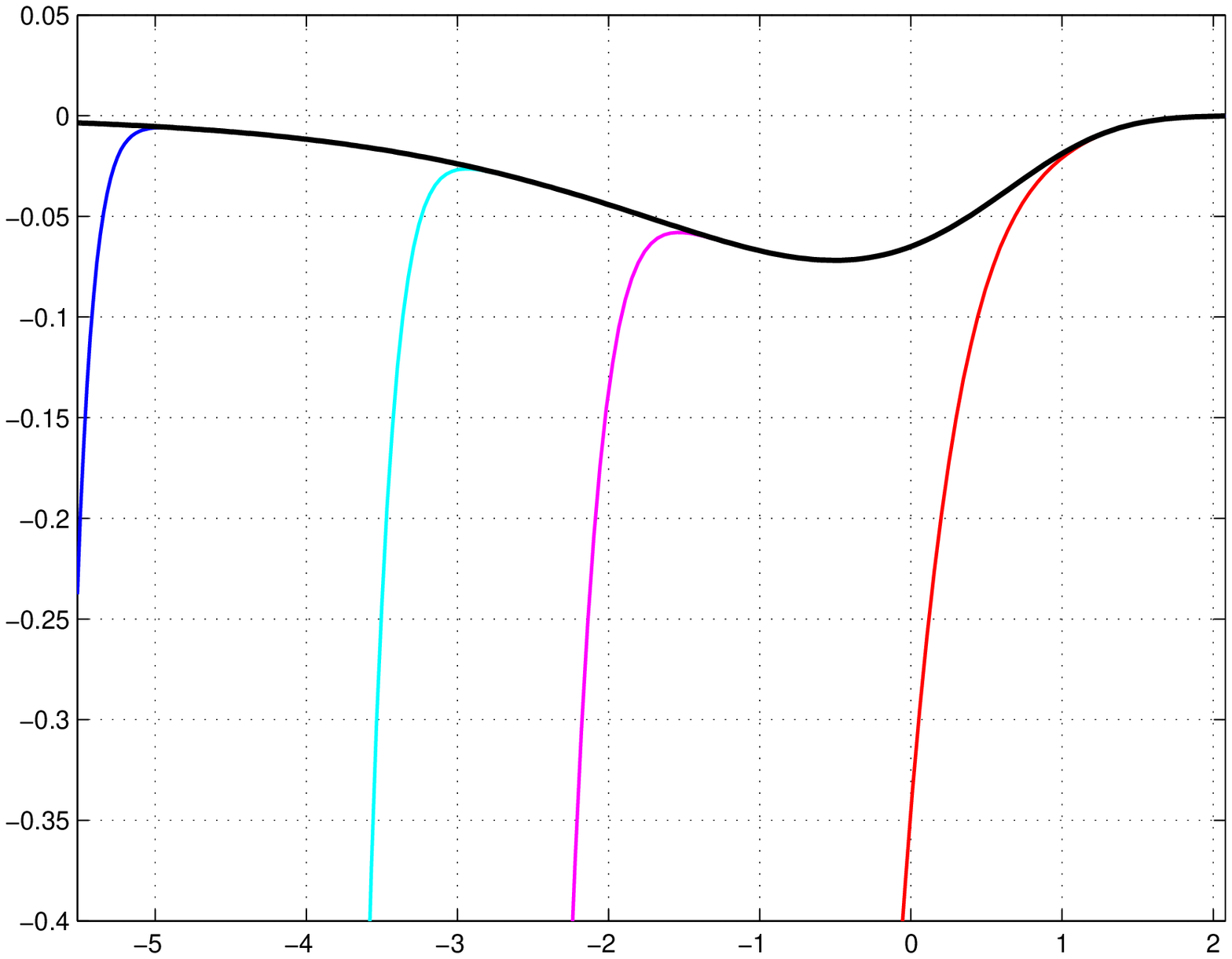}
&
\refstepcounter{figure}
\label{ifree}
\epsfxsize=.49\linewidth
\epsfbox{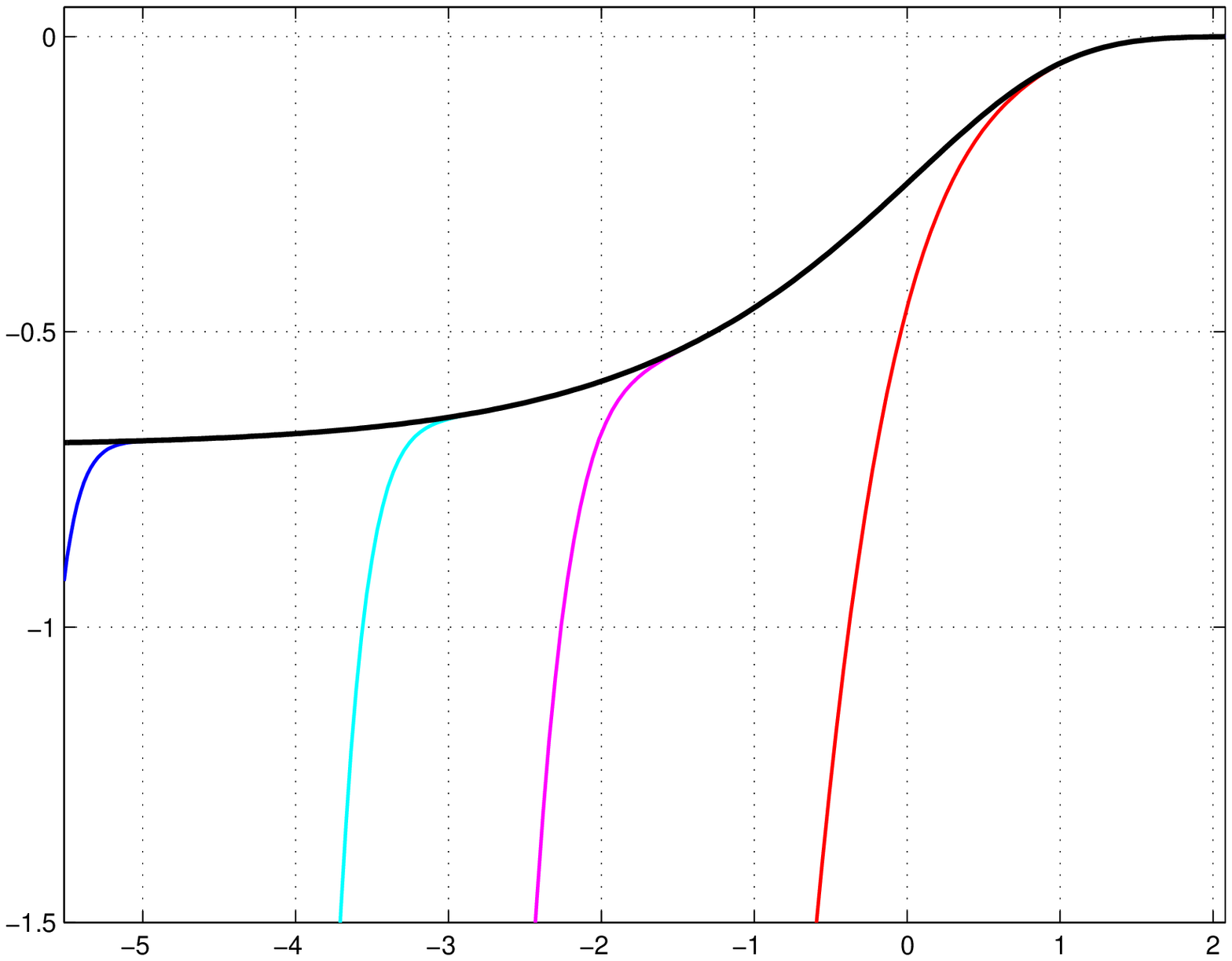}
\\
\parbox{.5\linewidth}{\small \raggedright
Figure \ref{ifix}:~$2\,\ln\, g_{\bf fixed}$  vs.\ $\ln(l)$
for Ising with fixed boundary conditions. {}From the bottom, the
lines represent 5, 100, 500, and 5000 particle contributions. The maximum
quantum number is 80 and $r=10$. The top line is
the exact result.} &
\parbox{.5\linewidth}
{\small \raggedright Figure \ref{ifree}:~$2\,\ln\, g_{\bf free} $
vs.\ $\ln(l) $ for Ising with free boundary conditions.
{}From the bottom, the lines represent 5, 100, 500, and 5000
particle contributions. The maximum quantum number is 80 and
$r=10$. The top line is the exact result.}
\end{array}
\]
\vskip 0.4cm

For interacting models a compact expression
such as (\ref{zz}) is not available, and one is forced to build the
partition function using the
LHS of (\ref{Pid}) directly. In figures \ref{ifix1} and
\ref{ifree1} we test this more general way to estimate a
$g$-function.
A similar  idea was first applied to the scaling Lee-Yang model in
\cite{Us1,Us3}; however in that case the  energy levels were
estimated using the BTCSA method~\cite{Us1}, rather than the Bethe
ansatz.

\[
\begin{array}{cc}
\refstepcounter{figure}
\label{ifix1}
\epsfxsize=.49\linewidth
\epsfbox{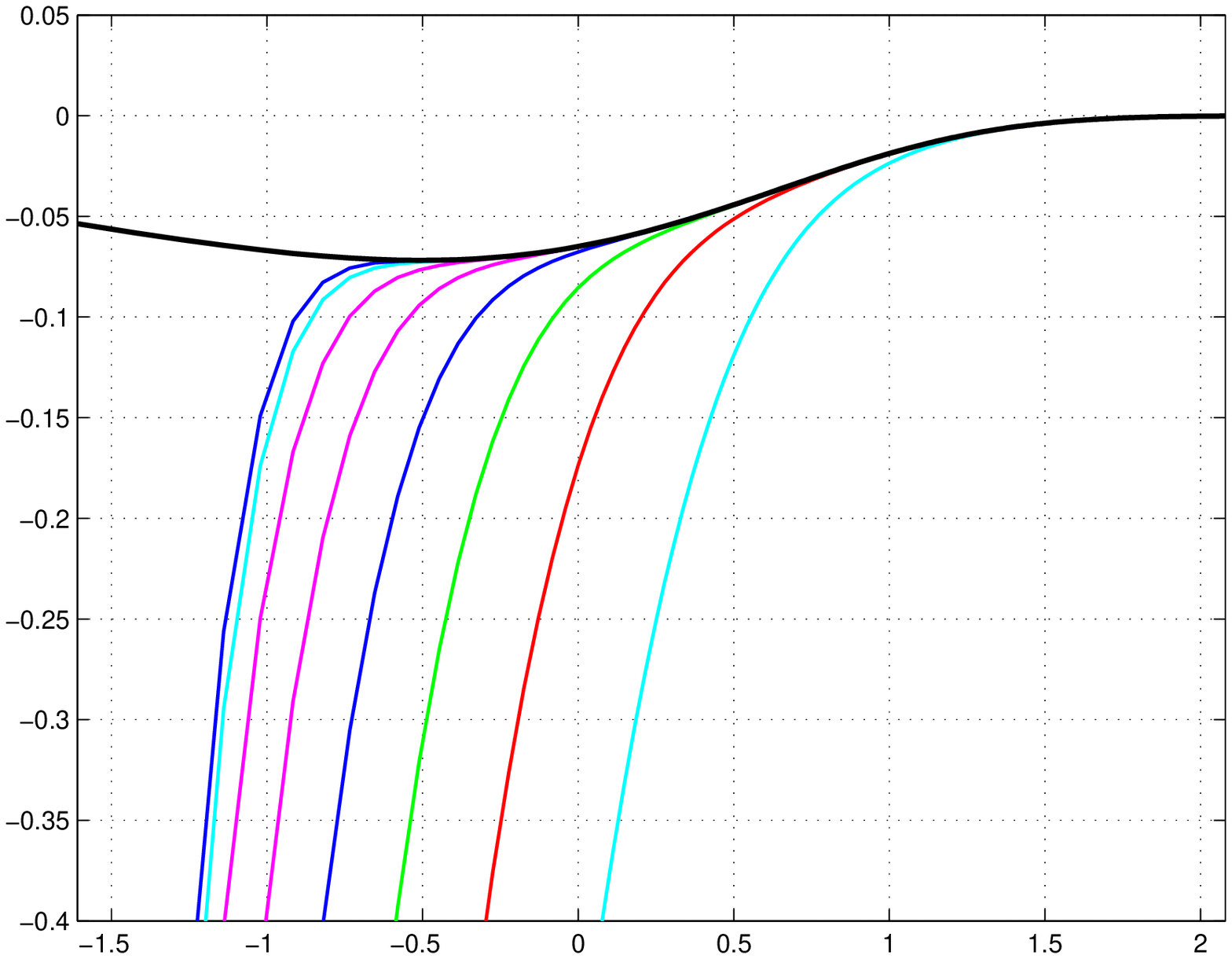}
&
\refstepcounter{figure}
\label{ifree1}
\epsfxsize=.49\linewidth
\epsfbox{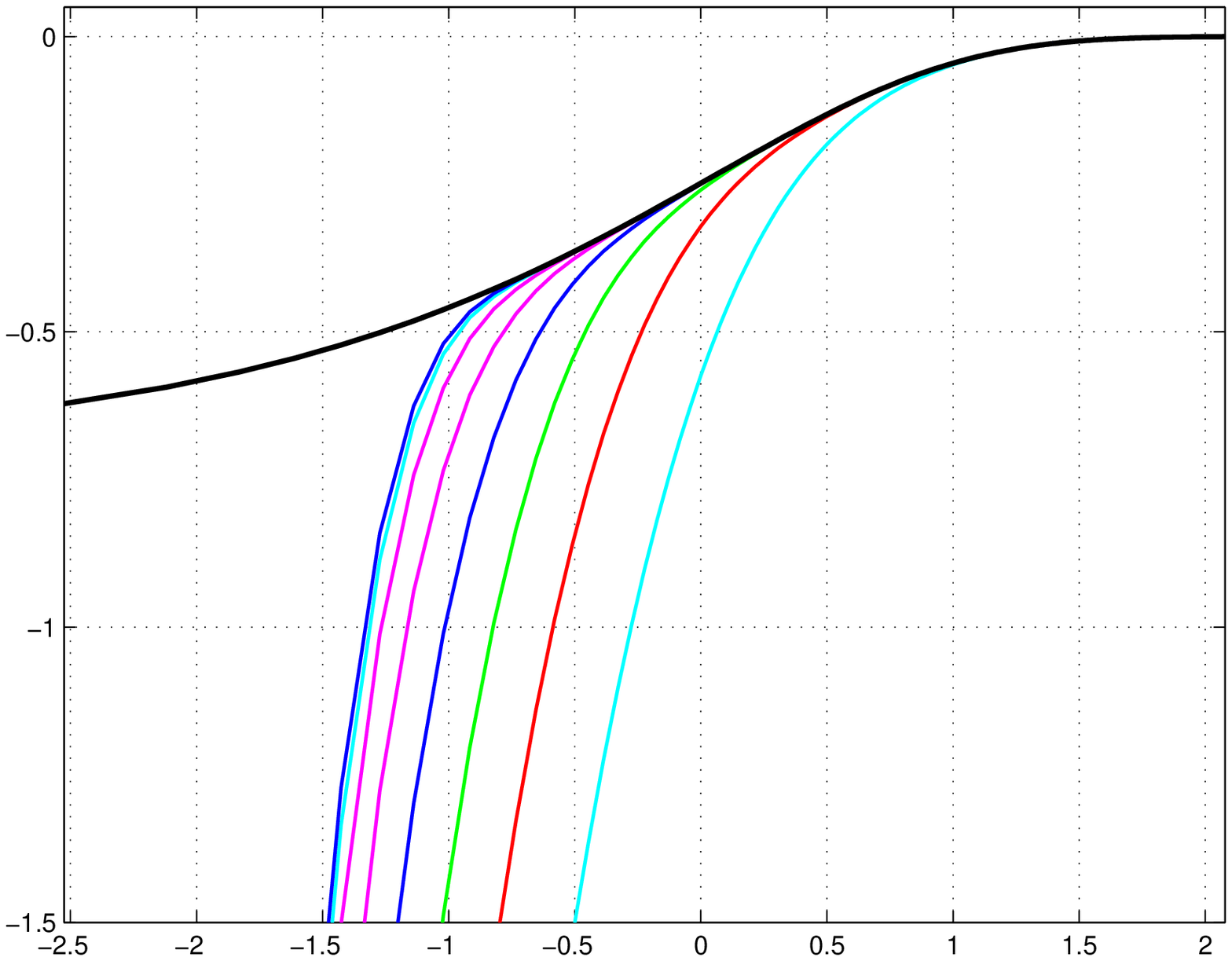}
\\
\parbox{.5\linewidth}{\small \raggedright
Figure \ref{ifix1}:~$2\,\ln\, g_{\bf fixed} $ vs.\
$\ln(l) $ for Ising with fixed boundary conditions. {}From the bottom,
the lines represent cluster contributions of $1,2,\dots,8$ particles.
The maximum quantum number 80 and $r=10$. The top line is
the exact result.} &
\parbox{.5\linewidth}
{\small \raggedright Figure \ref{ifree1}:~$2\,\ln\,
g_{\bf free}$ vs.\ $\ln(l) $ for Ising with free boundary conditions.
{}From the bottom, the lines represent cluster contributions of
$1,2,\dots,8$ particles. The maximum quantum number 80 and
$r=10$. The top line is the exact result.}
\end{array}
\]
\vskip 0.5cm
We see that it is hard to  get a good estimate of
the ultraviolet value of $\ln g$ from this form of the cluster expansion.
In section \ref{ly11} we shall solve this numerical problem
for the case of the scaling Lee-Yang model
by matching this numerical Bethe ansatz calculation with the
ultraviolet perturbed CFT results of~\cite{Us3}, while in section
\ref{IRLY} we shall develop a more analytical treatment.
\resection{Earlier proposals for $g$}
\label{earlier}
Consider a 1+1 dimensional integrable field theory with  entirely
diagonal  scattering and $N$ particle species.
According to the proposal of
\cite{LMSS} the  boundary entropy should be given by
an expression of the form
\bea
 \ln g_{\alpha}(l)
=   {1 \over 4} \sum_{a=1}^N \int_{\Rth} d \te \,
\Theta_a(\te)\ln(1+e^{-\varepsilon_a(\te)})
   \,,
\label{gf}
\eea
where the function  $\varepsilon_a(\te)$ is the solution of the
periodic-boundary-conditions TBA  (\ref{Ltbaa}),
and
\eq
\Theta_a(\te)=
\left(
   \phi_\alpha^{(a)}(\te)
 \;-\;
   2 \phi_{aa}(2\te)
 \,-\,
    \delta(\te)
\right)
\label{f1}
\en
with
\eq
\phi_\alpha^{(a)}(\theta)
= - {i \over \pi}  \frac{d}{d \te} \ln R_\alpha^{(a)}(\theta) ~,~~~
\phi_{ab}(\theta)
= - {i \over 2 \pi}  \frac{d}{d \te} \ln S_{ab}(\theta) ~.
\en
(Note, the normalisations of
$\phi_\alpha^{(a)}(\theta)$ and $\phi_{ab}(\theta)$
differ from those in \cite{Us3,RC} by factors
of $\pi$ and $2\pi$ respectively. This change is merely to simplify
some later formulae.)

However, the detailed analysis of \cite{Us3} showed that, for
non-zero values of the lightest bulk mass $M$, the resulting
$l$-dependence was incorrect, both in the total change in
$g_{\alpha}(l)$ between UV and IR, and in the behaviour of
the small-$l$ series expansion. On the other hand, the
predictions of (\ref{gf}) and (\ref{f1}) for
dependence of $g_{\alpha}(l)$
on the boundary parameters at fixed $l$, and also
for the  ratios of $g$-functions
${g_{\alpha}(l) / g_{\beta}(l)}$\,,
{\em were}\/
in very good agreement with  conformal perturbation theory and
the BTCSA. This suggested that the formulae should
be modified by some boundary condition independent extra terms, but
provided little clue as to what those extra terms should be.

Subsequently, it was proposed in \cite{RC}
that (\ref{f1}) should be replaced by
\eq
\Theta_a(\te)=
\left(\,
   \phi_\alpha^{(a)}(\te)
 \;-\;
   2 \phi_{aa}(2\te)
 \,-\,
    \phi_{aa}(\te)
\right)\,.
\label{f2}
\en
However, using results tabulated in \cite{Us3} it can be checked that
this modification does not cure the problems arising
in the bulk-massive case.
\resection{The scaling Lee-Yang model }
\label{ly11}
The spectrum  in the bulk scaling Lee-Yang theory consists of
a single particle species, with two-particle bulk scattering
amplitude
\cite{SmLY,CM}
\eq
  S(\te)
=  -(1)(2)
\;,\;\;
  (x)
=  {\sinh \lf( \fract{\te}{2} +\fract{i \pi x}{6} \ri)
     \over
    \sinh \lf( \fract{\te}{2} -\fract{i \pi x}{6} \ri)}~.
\en
When a boundary is present, two different types of boundary conditions
arise, which were labelled $\One$ and $\Phi(h)$
in \cite{Us1}. The corresponding reflection factors are
\eq
  R_{\Phi(h)}(\te)
= R_b(\te)
\;,\;\;\;\;
  R_\One(\te)
= R_0(\te)
\;,
\label{eq:ss3}
\en
where
\eq
h \sim   -|h_c|
\, \sin( (b+.5) \pi/5) M^{6/5}
~,~~~h_c= -0.6852899\dots
\en
is the coupling of the boundary field and
\eq
  R_b(\te)
= \lf(\fract{1}{2}\ri)\lf(\fract{3}{2}\ri)
  \lf(\fract{4}{2}\ri)^{-1}
  \lf( S(\te+i \pi \fract{b+3}{6})
  S(\te-i \pi \fract{b+3}{6}) \ri)^{-1}\,.
\label{bf}
\label{eq:ss2}
\en
We first use  the Bethe ansatz equation together with
(\ref{gexp1}) to obtain the $g$-function up to six-particle
contributions. The results are shown in figure \ref{figly} for the
boundary condition $\One$, and are compared with the ultraviolet
expansion obtained from (boundary) conformal perturbation theory
and the BTCSA \cite{Us3}:
\eq
 2 \ln g_\One(l)= { 1 \over 2} \ln\left( { \sqrt{5} -1
 \over 2 \sqrt{5}} \right) + 2 \,\frac{f_{\One}}{M}\,l
 + 2 \sum_{n=1}^4 d_n \left( \frac{l}{\kappa} \right)^{\!\frac{12}{5}n}+
 O(l^{12})\,,
 \label{UVp}
\en
 where $f_\One= {1 \over 4} (\sqrt{3}-1) M$,
 $d_1\approx -0.25312$, $d_2\approx 0.0775$, $d_3\approx -0.0360$,
 $d_4\approx 0.0195$, and $\kappa \approx 2.6429$.
\[
\begin{array}{c}
\refstepcounter{figure}
\label{figly}
\epsfxsize=.6\linewidth
\epsfbox{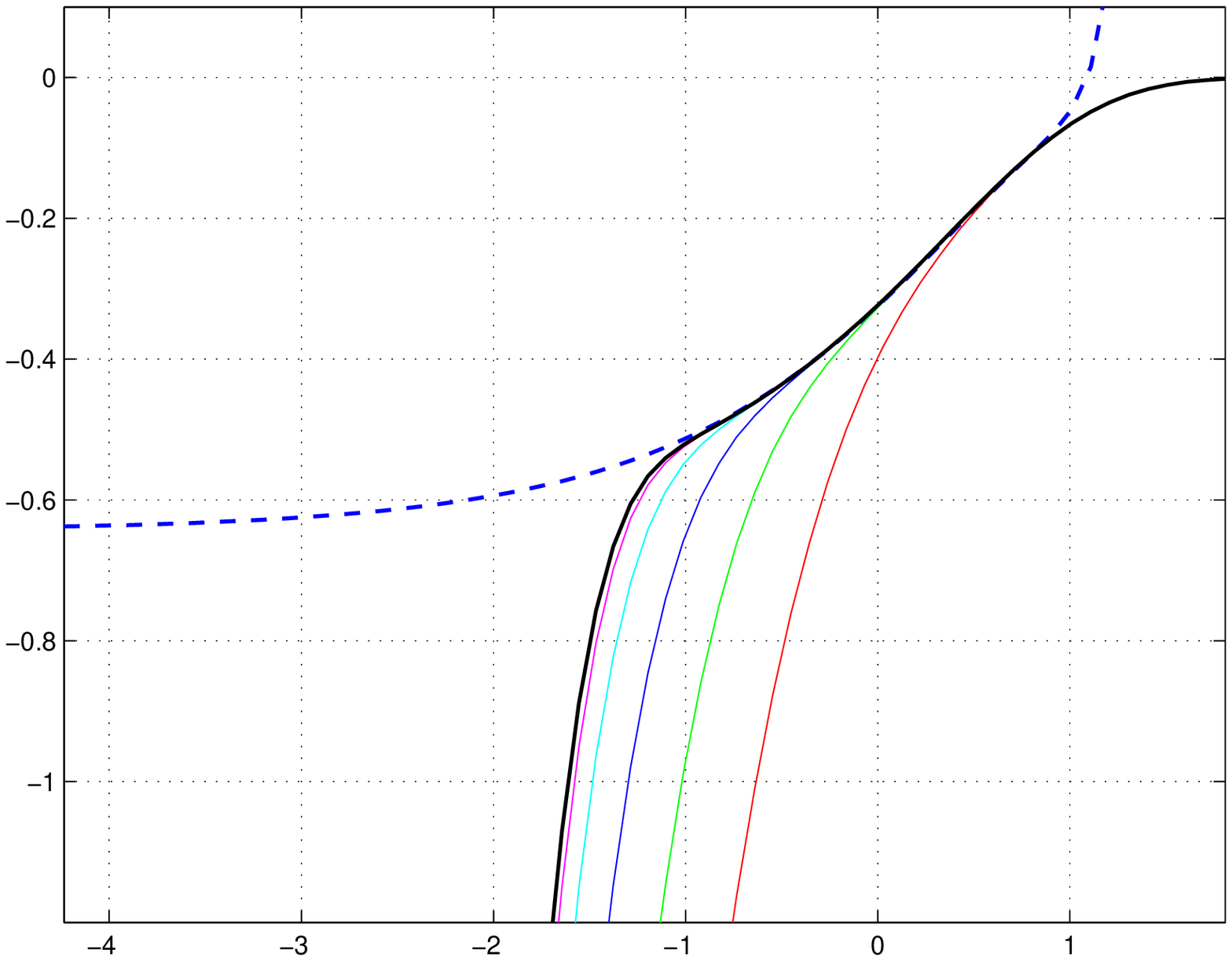}
\\
\parbox{0.95\linewidth}{\raggedright \small
Figure \ref{figly}: $2\,\ln\,g_\One$ vs.\
$\ln(l)$ with $(\One,\One)$ boundary conditions.
The dotted line is the $4^{\rm th}$
order result of \cite{Us3} and the solid
lines are BA results with $r=8$. The solid lines,
from the bottom, represent the $1,2,\cdots,6$ particle
contributions respectively. The maximum quantum number used was
80 (57 for the 6-particle
contribution).}
\end{array}
\]
As can be seen from the figure, the results from  the Bethe ansatz
and from
perturbed conformal field theory overlap over a
significant range of scales. This supports our hypothesis
that the two approaches
are describing the same function $g_\One(l)$, expanded about
either the IR or the UV.
\resection{Infrared  expansion for the Lee-Yang model }
\label{IRLY}
The purpose of this section is to develop an analytic technique to
check the earlier proposals described in
section~\ref{earlier} and at the same time
to give  hints about the appropriate modifications. The idea,
successfully applied above to the Ising model, is to start from the
Bethe ansatz
and to set up a cluster expansion by transforming the sums
into integrals as $R \rightarrow \infty$. The
method is quite  powerful, and already at first order
it confirms the questions raised in \cite{Us3} about
the proposals described in section~\ref{earlier}.
To simplify the discussion we shall only treat the $(\One, \One)$
boundary conditions directly. However,
this restriction is of no real significance
since the results in \cite{BLZ1,Us3,Us5} show that
\eq
{\cal G}_{\Phi(h)}^{(0)}(l)=Y \Big(i \pi {b+3 \over 6}
\Big) {\cal G}^{(0)}_\One(l)~,~~~Y(\te) = e^{\varepsilon(\te)}\,,
\label{gyy}
\en
where $\varepsilon(\te)$  is the  solution of the ground
state  TBA equation  with periodic boundary conditions.
\subsection{The one particle contribution}
\label{onep}
We start from the large-$R$ equation (\ref{gexp2}), truncated at the
one-particle level:
\eq
2\ln g_{\One} \sim R  \Bigl( E_0^{\rm circ}(M,L) - {\cal E}M^2L\Bigr)
+\ln \Big(1+ \sum_{n_1>0} e^{-l \cosh \te_1(n_1)} +\dots \Big)\,,
\label{lnz2}
\en
where the one-particle Bethe ansatz
essentially coincides with that for a
free particle,
\eq
r \sinh \te_1 -i  \ln R_{\One}(\te_1)  =  \pi n_1\,,
\en
with integer  $n_1$. Performing the continuous limit as
in section~\ref{ising} we find
\eq
P_1=\sum_{n_1>0} e^{-l \cosh \te_1} = {1 \over 2} \Big(
\sum_{n_1=-\infty}^{\infty} e^{-l \cosh \te_1} - e^{-l } \Big)
\longrightarrow { 1 \over 2}
\int_{\Rth} d \te\,
(J^{(1)}(\te)-\delta(\te))\,e^{-l \cosh \te}
%-{1 \over 2}   e^{-l} \,,
\en
where the Jacobian for the change of variable $n_1 \rightarrow
\te_1\equiv\theta$  is
\eq
J^{(1)}(\te)= {r \over \pi} \cosh\te+ \phi_{\One}(\te)\,.
\en
The cosh term cancels the leading part of the term linear in $R$ on the
RHS of (\ref{lnz2}), leaving
the  first contribution to $\ln g_\One$ as
\eq
2 \ln g_{\One}=  {1 \over 2} \int_{\Rth}  d \te
\left (\phi_{\One}(\te)-\delta(\te) \right)  e^{-l\cosh\te}+\dots\,.
\label{order1ch}
\en
Comparing this result with the proposals
 of section \ref{earlier}, we  conclude that
both are incorrect: in
particular no $\phi(2 \te)$ or
$\phi(\te)$ terms are involved in the
leading large $l$ asymptotic.

Next, we want to use this result to gain a hint as to how the
earlier proposals should be modified. However, the task  to
totally  or even  partially re-sum the cluster expansion directly
is, in principle, very hard. Our work is driven by the
extra assumption that  the final result should  depend, just
like the earlier `partially  correct'  proposals (\ref{f1})
and (\ref{f2}), on the bare single-particle energies only through
the TBA  pseudoenergies $\ep(\te)$. As will be reported
in more detail below, the consistency  of this assumption was
checked carefully up to four particles and confirmed, by a
more  superficial inspection, to all orders.

The attempt to find an
exact expression for $\ln g_\One$, therefore,  naturally starts from
\eq
\ln g_{\One} =  [\ln g_{\One}]^{(1)}_D+ \dots
\label{order1}
\en
where we have defined the `dressed' version of (\ref{order1ch}) to be
\eq
2 [\ln g_{\One}]^{(1)}_D= {1 \over 2}
 \int_{\Rth}  d \te
 \left (\phi_{\One}(\te) -   \delta(\te) \right)
 \ln(1+ e^{-\varepsilon(
 \te)}).
\label{order1D}
\en
Figure~\ref{figgt} gives some
initial numerical support for the conjecture.
\[
\begin{array}{c}
\refstepcounter{figure} \label{figgt}
\epsfxsize=.65\linewidth
\epsfbox{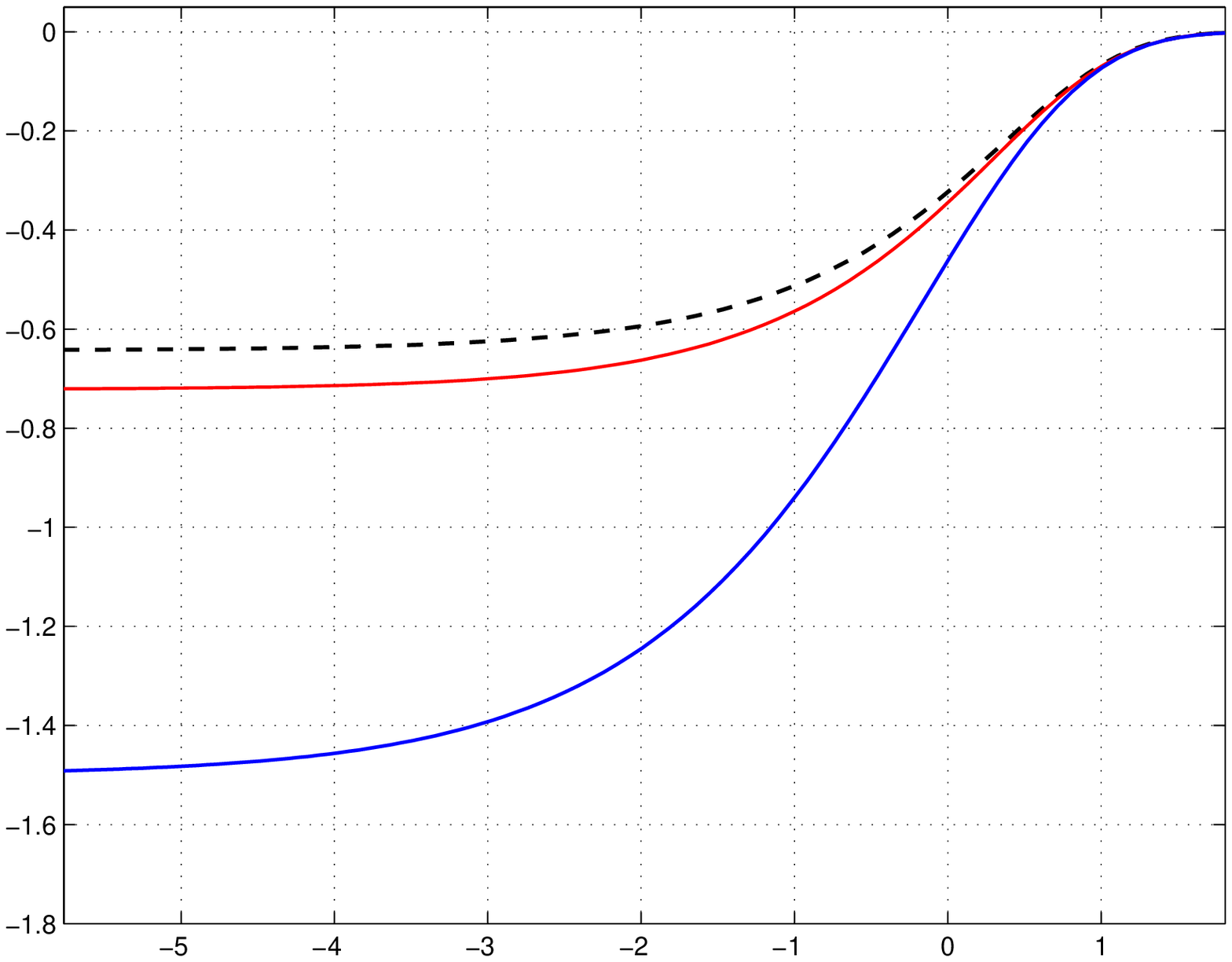} \\[2pt]
\parbox{.95\linewidth}{\raggedright \small
Figure \ref{figgt}:
$2\,\ln\,g_\One$ vs.\ $\ln(l)$. The top
dotted line is the `exact' result obtained by combining the CFT and BA
results from
figure
\ref{figly}. The solid lines are
obtained from (\ref{order1ch}) (bottom line) and
(\ref{order1}) (middle line).}
\end{array}
\]
Notice now that $[\ln g_{\One}]^{(1)}_D$ also
contains $n(>1)$-particle
 contributions. To see this at  second order, we expand
\eq
 \ln(1+e^{-\ep(\te)})~~~\hbox{as}~~~ e^{-\ep(\te)} -
{e^{-2 \ep(\te)} \over 2} + \dots\,,
\label{mod}
\en
and use   the exponential  of equation (\ref{Ltbaa})
expanded in terms of the bare particle energy $e(\te)=M
\cosh \te$
\eq
e^{-\ep(\te)} = e^{- l \cosh \te} \left (1+  \int_{\Rth}
 d \te'  \phi(\te- \te') e^{-l \cosh \te'} \right) + \dots
\en
to see that
\bea
&&
\!\!
\!\!
\!\!
\!\!
\!\!
2[\ln g_{\One}]^{(1)}_D =  {1 \over 2} \int_{\Rth} d \te
\, \left (\phi_{\One}(\te) -  \delta(\te) \right)  e^{-l \cosh \te}
-{1 \over 2}
\int_{\Rth}  d \te \,  \phi(\te)  e^{- l \cosh \te - l}
+{ 1 \over 4} e^{ -2l}
\qquad
\qquad
 \nn \\[3pt]
 &&
\!\!
\!\!
{}- { 1\over 4} \int_{\Rth} d \te \,\phi_{\One}(\te)
e^{- 2l \cosh \te} + { 1 \over 2} \int_{\Rth^2} d \te_1  d \te_2
\, \phi_{\One}(\te_1)
\phi(\te_1-\te_2)  e^{- l \cosh \te_1 - l \cosh \te_2}\,+ \dots
\qquad
\quad
\label{onef}
\eea
The aim of the analysis in the next sections is to justify
the replacement of $e^{-l\cosh \te}$ by
$\ln(1+e^{-\varepsilon(\te)})$  and is also  to find
some hints as to the origin and form of the further correction
terms in (\ref{order1}).

\subsection{Two and three particle contributions }
\label{twop}
We start again from (\ref{gexp2}), this time keeping both one and
two particle contributions:
\bea
2\ln g_{\One}&=&
R  \Bigl( E_0^{\rm circ}(M,L) - {\cal E}M^2L\Bigr) \nn\\[3pt]
&&{~}+~
\ln \Big (1+ \sum_{n_1>0} e^{-l \cosh \te_1} +
\sum_{n_1>0} \sum_{n_2>n_1}
e^{-l\cosh \te_1-l \cosh \te_2} +\dots \Big )\,,~~~~
\label{lnz2a}
\eea
where the two-particle-state momenta $(\theta_1,\theta_2)$ are
related to their quantum numbers $(n_1,n_2)$ via the Bethe ansatz
equations
\bea
 {r \over \pi} \sinh \te_1 -{i \over \pi}  \ln R_{\One}(\te_1)-
{i \over 2 \pi} \ln S(\te_1-\te_2) S(\te_1+\te_2)
&=& n_1\,; \nn \\
{r \over \pi} \sinh \te_2 -{ i \over \pi} \ln R_{\One}(\te_2)- { i
\over 2 \pi} \ln S(\te_2-\te_1) S(\te_2+\te_1)
&=& n_2\,.~~~~~~~~
\label{baetwo}
\eea
The new piece in (\ref{lnz2a}) can be  written as
\bea
\!\!\!
\!\!\!
P_2 &=& \sum_{n_1>0} \sum_{n_2>n_1} e^{-l \cosh \te_1- l\cosh \te_2}
\equiv { 1 \over 8} \sum_{n_1=-\infty}^{\infty}
 \sum_{n_2=-\infty}^{\infty} e^{-l \cosh \te_1- l\cosh
 \te_2} \nn \\
&&{}-~ {1 \over 4} \sum_{n_1=-\infty}^{\infty} e^{-l \cosh \te_1- l}
-{1 \over 4}
\sum_{n_1=-\infty}^{\infty} e^{-2 l \cosh \te_1}
+ {3 \over 8} e^{-2 l}\,.
\label{secor}
\eea
As $R \rightarrow \infty$
the continuous limit can be taken as:
\bea
\ds{ \sum_{n_1>0} \sum_{n_2>n_1} e^{-l \cosh \te_1- l\cosh \te_2}
\longrightarrow { 1 \over 8} \int_{\Rth^2}  d \te_1 d \te_2
   J_1^{(2)}(\te_1,\te_2)
e^{-l \cosh \te_1 -l\cosh \te_2}}  \nn \\
\ds{ - {1
\over 4} \int_{\Rth} d \te    J_2^{(2)}(\te)  e^{-l \cosh \te-l}
-{1 \over 4}
\int_{\Rth} d \te  J_3^{(2)}(\te) e^{-2 l \cosh \te}
+ {3 \over 8} e^{-2 l}}
\eea
The Jacobians $J_1^{(2)}(\te_1,\te_2)$, $J_2^{(2)}(\te)$ and
$J_3^{(2)}(\te)$  can be calculated
from the Bethe ansatz equations (\ref{baetwo})
as before. Notice that the
correct subtractions of the excluded  contributions (those excluded by
the statistics) are crucial to get the corresponding
Jacobians $J_2^{(2)}(\te)$ and $J_3^{(2)}(\te)$: one has to
take the derivatives only after the forbidden values of the
quantum numbers  $n_1$ and $n_2$ are fixed.
 Although we have performed the calculation in full
and checked the consistent cancellations of the $r$ (strip size)
dependent  parts, for brevity we shall concentrate on the
subleading, $r$-independent, parts $j_1^{(2)}$, $j_2^{(2)}$, $j_3^{(2)}$:
\bea
j_1^{(2)}(\te_1,\te_2)
&=& \phi_{\One}(\te_1)\phi_{\One}(\te_2)
+2 \phi(\te_1-\te_2) \phi_{\One}(\te_2)+
2 \phi(\te_1-\te_2) \phi_{\One}(\te_1) \nn \\
&&{}+\, 4 \phi(\te_1+\te_2)  \phi(\te_1-\te_2)~,  \nn \\
j_2^{(2)}(\te) &=& \phi_{\One}(\te)  +2\phi(\te) ~,
\quad j_3^{(2)}(\te) = \phi_{\One}(\te)  +2 \phi(2 \te)
\,.
\eea
Expanding the logarithm  in (\ref{lnz2a}),
we have at second order
\eq
\ln(1+P_1+P_2+\dots)  = P_1+ \left( P_2- {P_1^2 \over 2} \right)+\dots\,.
\en
The up-to second order  $2 \ln g_{\One}$ contains seven distinct
contributions, the first five  coinciding  with those
written explicitly on the RHS of (\ref{onef}) and corresponding  to
the up-to-two particle expansion of $2[\ln
g_\One]^{(1)}_D$. (This confirms  the correctness of
the dressed formulae (\ref{order1}, \ref{order1D}) up to this point).
There are also two genuinely new terms, and we find:
\bea
2 \ln g_{\One} &=& 2[\ln g_\One]^{(1)}_D  + { 1 \over 2} \int_{\Rth^2} d\te_1   d \te_2 \, \phi(\te_1 +\te_2)
\phi(\te_1-\te_2) e^{-l
\cosh
\te_1 -l
\cosh
\te_2} \nn \\
&-&{ 1 \over 2} \int_{\Rth} d \te \,   \phi( 2 \te)  e^{-2 l \cosh \te}+\dots~,
\label{newone}
\eea
The final step is to iterate the dressing procedure, though in
a modified form, by replacing   $l\cosh \te$ with
$\varepsilon(\te)$ and writing $\ln g_{\One} =
 [\ln g_{\One}]_D^{(1)}+  [\ln g_{\One}]_D^{(2)}+\dots
$
with
\eq
2 [\ln g_{\One}]_D^{(2)} =  { 1 \over 2} \int_{\Rth^2}  d\te_1  d \te_2 \,
\phi(\te_1 +\te_2)
\phi(\te_1-\te_2) e^{-\ep(\te_1) - \ep( \te_2)}
- { 1 \over 2} \int_{\Rth} d \te \, \phi( 2 \te) e^{-2 \ep( \te)}~. \nn
\en
Again this dressing prescription can be justified retrospectively by
testing at  third and higher  order. The third-order  result
turns out to support the assumption, and gives
a genuinely  new type of  correction  to
$2 \ln g_\One$,  independent of $\phi_\One(\te)$:
\bea
2 \ln g_\One &=&  2 [\ln g_{\One}]_D^{(1)}+ 2[\ln g_{\One}]_D^{(2)}
+
 {2 \over 3 } \int_{\Rth} d \te\,   \phi(2 \te) e^{-3 l \cosh \te} \nn \\
&+&  {1
\over 3} \int_{\Rth^3} d \te_1  d \te_2
d \te_3 \, \phi(\te_1+\te_2)  \phi(\te_2 -\te_3)
 \phi(\te_3 - \te_1) e^{-l \cosh \te_1 -l \cosh
\te_2 -l \cosh \te_3}
\nn \\
&-&  \int_{\Rth^2} d \te_1 d \te_2 \, \phi(\te_1+\te_2)
\phi(\te_1-\te_2)  e^{-l \cosh \te_1 - 2 l \cosh \te_2}+ \dots\,.
\label{newtwo}
\eea
\subsection{The exact result}
\label{exact}
To go further in the expansion becomes increasingly difficult, due
to higher number of Jacobians and the huge number of terms
contributing to a  single Jacobian. However we managed to
complete the analysis up to  four particles and to perform a more
superficial inspection at higher orders. The following results
were deduced: at each order there is always a new
contribution of the form
\eq
C_n=
 {1\over n}
 \int_{\Rth^n}  d \te_1  \dots d \te_n\,
 \phi(\te_1+\te_2)\phi(\te_2-\te_3)\dots
\phi(\te_{n}-\te_1){e^{-l \cosh\te_1}} \dots {e^{-l \cosh \te_n}}
\,,
\en
when the lower order terms
$C_{2}, \dots, C_{n-1}$ are  corrected  according to the rule
\eq
e^{-l \cosh \te} \rightarrow {1 \over 1+ e^{\ep(\te)}}\,.
\en
An additional term contains a single integration over the function
$\phi(2 \te)$, as already seen in (\ref{newone}) and
(\ref{newtwo}).
To treat these terms
is trickier, but after a few attempts we convinced ourselves
that the correct procedure is always to replace $l\cosh\theta$ with
$\ep(\theta)$, and then to resum, as:
\eq
\int_{\Rth} d \te \, \phi( 2 \te)
\left(
- { 1 \over 2} e^{-2l\cosh\te}
+\dots
\right)
\rightarrow -\int_{\Rth} d
\te  \, \phi(2 \te) \Big( \ln(1+e^{-\ep(\te)})-
{1 \over 1+e^{\ep(\te)}} \Big)\,.
\label{subs}
\en
We  finally   arrive at
\bea
&&2\ln g_\One(l)= {1 \over 2}
   \int_{\Rth}
d \te \,
 (\phi_\One(\te) -  \delta(\te) - 2\phi(2 \te) )
 \ln(1+ e^{-\varepsilon(
 \te)})
  \nn \\
&+&
\sum_{n=1}^{\infty}
{1\over n}
\int_{\Rth^n}  {d \te_1  \over 1+e^{\ep(\te_1)}} \dots
{d \te_n   \over 1+e^{\ep(\te_n)}}\,
\phi(\te_1+\te_2)\phi(\te_2-\te_3)\dots \phi(\te_{n}-\te_{n+1})\,,~~~~~~~~
\label{finalc}
\eea
with $\te_{n+1}=\te_1$ \footnote{Notice the
sequence of
signs $+,-,-,\dots,-$ in the arguments of the $\phi$s in
(\ref{finalc}), that the number of $\phi$ factors in the
$n^{\rm th}$  correction term is $n$ and that according to
(\ref{subs}) at $n=1$ only $\phi(2 \te)$ survives.}.
It is not hard to check that this formula
is consistent with the one-, two-, and three-
particle results described above.
Another simple (but nontrivial)
check can be performed in the ultraviolet, as
\eq
e^{\ep(\te)} \rightarrow e^{\ep_0}={
\sqrt{5}+1
\over 2}\,,~~~~\int_{\Rth} d\te\, \phi_{\One}(\te)=-2\,,
\en
\eq
\int_{\Rth} d\te\, \phi(\te)=-1\,,~~~~ \int_{\Rth^n}
d \te_1 \dots  d \te_n \,
 \phi(\te_1+\te_2)\phi(\te_2-\te_3)\dots
 \phi(\te_{n}-\te_{1})={(-1)^n \over 2}\,,
\en
 and (\ref{finalc}) reduces to
\bea
- \ln(1+e^{-\ep_0}) &+&
\sum_{n=1}^{\infty} {(-1)^n \over 2 n}
\Big({1 \over 1+ e^{\ep_0}} \Big )^n= - \ln(1+e^{-\ep_0}) -{1 \over 2}
\ln \Big (1+ {1 \over 1+e^{\ep_0}} \Big) \nn \\
 &=&\frac12 \ln \Big(\frac{\sqrt5-1}{2\sqrt5} \Big)\,,
\label{losum}
\eea
which is the expected $l=0$ value for
$2 \ln g_{\One}$ given in (\ref{UVp}).

The first term on the RHS of (\ref{finalc}) coincides with the proposal
of \cite{LMSS}, repeated in equations (\ref{gf}) and (\ref{f1}) above,
while the remaining parts constitute a boundary-condition independent
correction, which only comes nontrivially into play when the bulk theory
is massive. As mentioned at the end of section \ref{earlier}, this was
only to be expected given the results of \cite{Us3}, but it is
nevertheless satisfying that the already-verified portions of
the earlier results
have been recovered by this rather different route.

In figure \ref{figg_3_log} the result from equation (\ref{finalc})
is compared with the `exact' result obtained
in section~\ref{ly11}
by combining
UV perturbed CFT results
with the IR cluster expansion from the Bethe ansatz.
We see that keeping  only
the first three terms in the series already gives a very good agreement
with the exact result: in the UV the agreement is within
about $0.3\% $.
\[
\begin{array}{c}
\refstepcounter{figure} \label{figg_3_log}
\epsfxsize=.60\linewidth \epsfbox{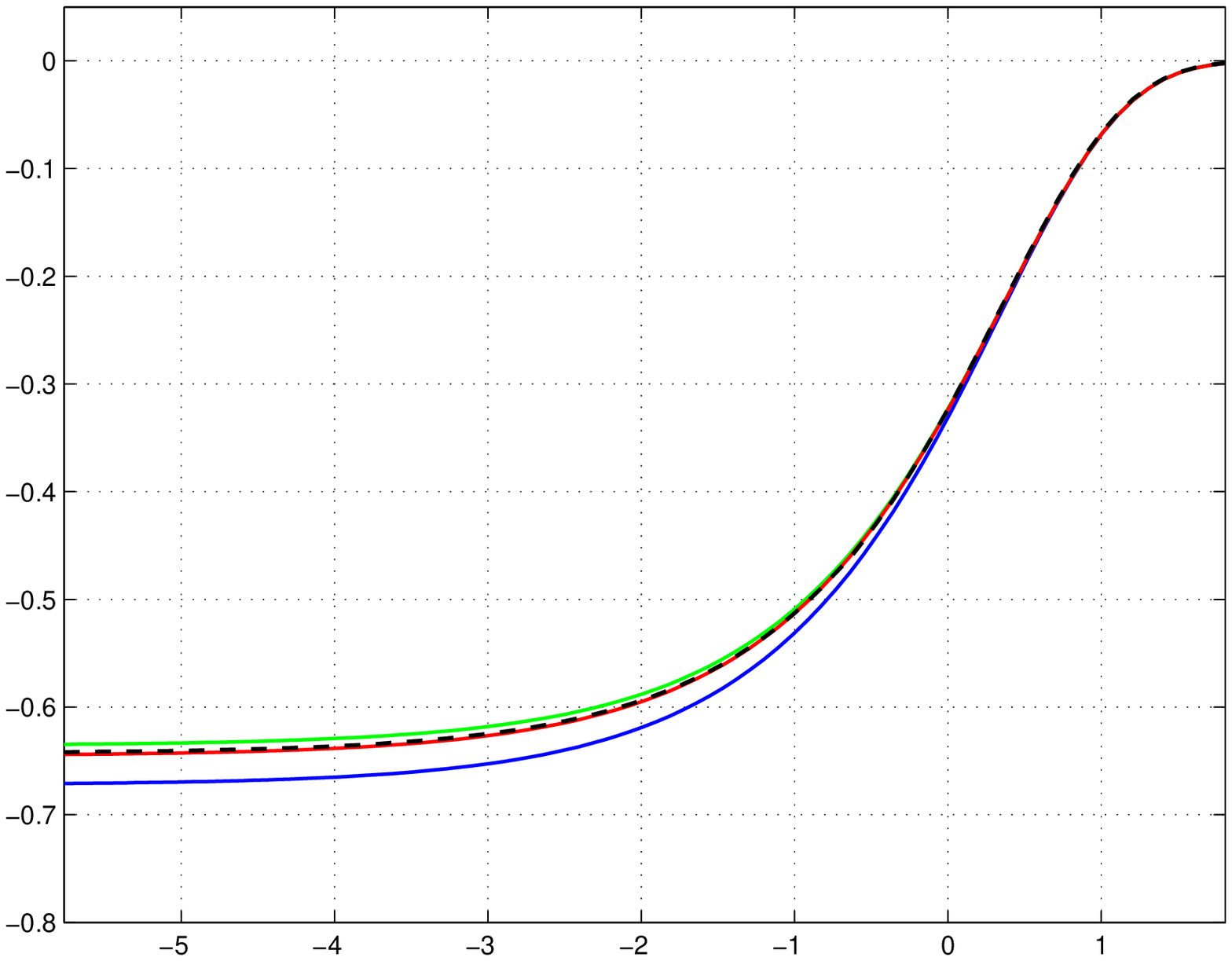}
\\[2pt]
\parbox{.95\linewidth}{\raggedright\small
Figure \ref{figg_3_log}:
 $2\,\ln\,g_\One$ vs.\ $\ln(l) $.
The dotted line corresponds to the `exact' contribution.
The bottom  solid line represents
the RHS of (\ref{finalc}) with the sum truncated at the first term.
The solid line just above the dotted line
truncates the sum at the doubly-integrated term,
and the line just below the dotted line
is the total contribution up to the triple integral.}
\end{array}
\]
We also used (\ref{finalc}) to make a numerical estimate of
the coefficients of the ultraviolet expansion
of $\ln g_{\One}(l)$:
\bea
&&\hskip -1.45truecm
\ln g_{\One}(l) \sim  -0.3214826953191634 +
0.483692443734693\,x^{\frac{5}{12}}
-0.253117570\,x \nn \\
  &&\hskip-.2truecm{}
    +0.0781176\,x^{2}
    -0.037284\,x^{3} +
     0.02042\,x^{4} -
     0.0120\,x^{5} +
     0.0074\,x^{6} +
     \dots\quad
\label{fit1}
\eea
with $x= (l/\kappa)^{12/5}$.  This can be compared with the result of \cite{Us3}:
\bea
&&\hskip -2.25truecm
\ln g_{\One}(l) \sim -0.3214826953191634 +
0.4836924437346968\,x^{\frac{5}{12}}
 - 0.253117581\,x  \nn \\
  &&\hskip-0.9truecm{}
  + 0.0775\,x^{2} -0.036\,x^{3} +
  0.0195\,x^{4}+\dots
\label{fit2}
\eea
(Exact expressions for the first three coefficients in
(\ref{fit2}) were found in \cite{Us3}; here we only
quote sufficiently-many digits to enable the numerical errors in
(\ref{fit1}) to be assessed.)

By inspection, it is straightforward to generalize (\ref{finalc})
to more general theories with  $N$ particle species and
entirely diagonal scattering and reflection  matrices. The
proposal is
\bea
2\ln g_\alpha(l)&=& {1 \over 2}
 \sum_{a=1}^{N} \int_{\Rth}
d \te
 \left (\phi_\alpha^{(a)}(\te) -  \delta(\te) - 2\phi_{aa}(2 \te) \right)
 \ln(1+ e^{-\varepsilon_a(
 \te)})
 \nn \\
&+&
\sum_{n=1}^{\infty}
\sum_{a_1\dots a_n=1 }^N
 {1\over n}
 \int_{\Rth^n}  {d \te_1   \over 1+e^{\ep_{a_1}(\te_1)}}  \dots  {d \te_n  \over 1+e^{\ep_{a_n}(\te_n)}} \times\nn \\
&&
 \left(
 \phi_{a_1 a_2}(\te_1+\te_2)\phi_{a_2 a_3}(\te_2-\te_3)
 \dots \phi_{a_{n} a_{n+1}}(\te_{n}-\te_{n+1})
 \right)\,,
\label{finald}
\eea
where  $\te_{n+1}=\te_1$, $a_{n+1}=a_1$, while $\phi_{ab}(\te)$ and $\phi_{\alpha}^{(a)}$ are defined in
section~\ref{earlier}. Note, though, that in some circumstances extra terms may be needed for the correct
analytic continuation of the integrals, as discussed in section 4.3.1 of \cite{Us3}.

Before concluding this section
we would like to mention that while this project was in progress
and some of the analytic
results already obtained as they  are written here,
a  preprint by Woynarovich appeared
\cite{Woy}.
 Our result
(\ref{finalc}) is similar in form
to the expression proposed by Woynarovich
for the $O(1)$ corrections to the free energy
for a one dimensional Bose gas with repulsive
$\delta$-function interaction. However, there is also a
major difference. The string of kernels
\eq
\phi(\te_1+\te_2)\phi(\te_2-\te_3)
\dots \phi(\te_{n}-\te_{n+1})\,,
\label{string1}
\en
in our (\ref{finalc}) is replaced by a string of the form
\eq
 \psi(\te_1,\te_2) \psi(\te_2,\te_3)\dots \psi(\te_{n},\te_{n+1})\,,
\label{string2}
\en
with $2 \psi(k_1,k_2) =\phi(k_1+k_2)+
\phi(k_1-k_2) \equiv   \overline K(k_1,k_2)$ in eq.~(3.28) of
\cite{Woy}~\footnote{Notice that the kernels in the two
cases are actually different, but the derivation in
\cite{Woy} is quite general, and the result is
independent of the precise functional form of the kernel.}.
That terms of the type (\ref{string1}) appear in our formulae
and not the expression (\ref{string2}) is unmistakably emerging
from the Jacobians for the change of variable $\{n_i\}
\rightarrow
\{\te_i \}$ and from their definitions as determinants. Woynarovich
obtained his result by a
calculation of the next-to-leading
contributions to the free energy, evaluating corrections to the
standard saddle point result. Such a direct computation
would be a highly
desirable alternative to the more indirect approach taken in this
paper. Unfortunately, as stated in the
paragraph after eq.~(5.8) in section V of \cite{Woy},
for the field theory case
the result of \cite{Woy} is
divergent in the ultraviolet, $R=1/T
\rightarrow 0$, limit. This rules out the possibility of its
consistent agreement with a $g$-function defined in
(perturbed) conformal field theory, of the sort studied in
\cite{LMSS,Us3,RC} and this paper.
Nevertheless, the mathematical similarity between the final
outcomes is  striking and deserves further investigation.
To make a more precise comparison note that
\eq
2 \log g_{\One}^{\mbox{(this~paper)}}\leftrightarrow
\left[ - T^{-1} ( \Delta F + \phi_0 +\phi_L) + \overline{\Delta S}~
\right]^{\mbox{(ref.\,\cite{Woy})}}
\en
and that  $-T^{-1} ( \Delta F + \phi_0 +\phi_L)$ matches
the first, single-integral,
term on the RHS of (\ref{finalc}). $\overline{\Delta
S}$ should then be compared with   the  infinite series
in (\ref{finalc}). In eq.~(5.8) of \cite{Woy} Woynoarovich notes that
his $\overline{\Delta S}$ can be written as a sum of two contributions:
an UV convergent part corresponding  to $1/2$  of our infinite series,
plus an UV divergent term which has no counterpart in (\ref{finalc}).
Thus, in spite of the apparent similarities between our results
and those of \cite{Woy}, there are also important discrepancies, which
we are currently unable to    resolve
physically.

\resection{Conclusions}
\label{conclusions}
This paper concerned the off-critical version of the boundary entropy $g$ as defined in field theory via the identity
(\ref{Pid}). It was shown numerically that the asymptotic infrared expansion for $\ln g$, obtained using the Bethe ansatz,
matches UV results from conformal perturbation theory and the BTCSA at intermediate scales. This was a crucial step in the
analysis, because it meant that these two alternative  definitions are equivalent, and it opened up the interesting
possibility of deriving an exact expression for the conformally-perturbed $g$-function by using the Bethe ansatz technique.
The first step toward this was to give the exact prescription, as the width of the strip $R$ tends to infinity, to
transform sums over the quantum numbers into integrals on rapidity variables.  The idea is that the subleading
$R$-independent terms in this expansion should build up to form the boundary entropy $g$. A careful inspection of this
expansion, motivated by  the plausible assumption that the final result should depend on the bare single-particle energies
only through  their dressed versions, i.e.\ through the TBA pseudoenergies $\ep(\te)$, led to a partial resummation with
corrections written purely in terms of $\ep(\te)$. The expressions for $\ln g$ written in (\ref{finalc}) and in
(\ref{finald})  are the main new results of the paper. Equation (\ref{finalc}) was carefully checked against results
obtained using a combination of conformal perturbation theory and the Bethe ansatz. The agreement was extremely good, and
showed that the series is rapidly convergent even in the ultraviolet region (see equations (\ref{losum}), (\ref{fit1}) and
(\ref{fit2})). Although it relied at various points on conjectures, we would also like to stress that our derivation
avoided some of the pitfalls that potentially afflict more direct computations of the $g$-function: by working in the
$l\to\infty$ limit, we always dealt with states in which all constituent particles were well-separated, and so the accuracy
of the Bethe ansatz wavefunctions for high particle density was not an issue.

There are many open problems related to this project,
the first being that the method proposed, for all its virtues,
is not particularly elegant and a direct approach
would be desirable for the generalization to
more complicated models. It would also
be interesting  to study the corresponding quantities in
theories with  non-diagonal scattering and in
systems with massless excitations in the  bulk~\cite{LSS}.\\

\noindent
{\bf Note added:}
\medskip

There is a numerical error in the third term of the expansion for $\ln g_{\One}$ given in eq. (5.29)
above, which was pointed out to us by Aliosha Zamolodchikov. This is due to an inaccuracy in Mathematica's
evaluation of the generalised hypergeometric function ${}_3F_2$\,, which arises in formula (3.14) of
ref.~[13] for the associated quantity $I_2$. In fact, Aliosha Zamolodchikov has found a simplified
expression for $I_2$\,, as follows:
\begin{eqnarray}
I_2&=&
\frac{5}{8}\log 5-\frac{5\sqrt{5}}{8}\log
\frac{\sqrt{5}+1}{\sqrt{5}-1}+\frac \pi 4\cot \frac{2\pi}{5}
\nonumber\\
&=&-0.08393791256821845466150\dots
\end{eqnarray}
(this contrasts with the value $-0.083937990\dots$ quoted in ref. [13]). Accordingly, the prediction
(5.29) above should be corrected to
\begin{eqnarray}
&&\hskip -2.25truecm
\ln g_{\One}(l) \sim -0.3214826953191634 +
0.4836924437346968\,x^{\frac{5}{12}}
  \nonumber \\
  &&\hskip-0.9truecm{}
 - 0.253117570093371858\,x
  + 0.0775\,x^{2} -0.036\,x^{3} +
  0.0195\,x^{4}+\dots
\end{eqnarray}
with an even better match to (5.28). We would like to thank Aliosha Zamolodchikov for discussions of this
point.

\medskip
\noindent
{\bf Acknowledgements}
\medskip

\noindent
We are grateful to  Z.~Bajnok, L.~Palla, G.M.T.~Watts and Al.B.~Zamolodchikov
for many useful  discussions. We would also
like to thank F.~Woynarovich for sending us an early version of~\cite{Woy}.

DF thanks the Leverhulme Trust (grant F/00224/G) for
a fellowship, and RT  thanks the EPSRC for an Advanced Fellowship.

This work was partly supported by the EC
network ``EUCLID", contract number HPRN-CT-2002-00325, and partly by
a NATO grant PST.CLG.980424. PED was also
supported in part by JSPS/Royal Society grant and by the
Daiwa Foundation, and the work by CR was partly supported by the Korea
Research Foundation 2002-070-C00025.

 PED, PED-DF-RT, and CR, thank
SPhT Saclay,  APCTP  Korea, and the Universities
of Durham and Torino, respectively,
 for hospitality while this project was in progress.
\appendix
\resection{TBA and BA in purely elastic scattering models}
\label{TBABA}
In this appendix we summarise the  equations relevant to our analysis.
\subsection{Periodic boundary conditions:}
\label{TBAC}
The thermodynamic Bethe ansatz equations are
\cite{AlZam1}
\eq
  \epsilon_a(\theta)
=
   M_a L \cosh\te
 - \sum_{b=1}^{N} \int_{\Rth}
d \te'  \,
   \phi_{ab}(\te - \te')\,
   L_b(\te')\,, ~~~~~~(a=1,\dots, N).
\label{Ltbaa}
\en
The ground state energy on a circle is expressed in terms
of the functions $L_a(\te)=\ln(1+e^{-\ep_a(\te)})$ as
\eq
   E_0^{\rm circ}(M,L) = -\sum_{a=1}^{N}
   \int_{\Rth} {d \te \over 2\pi} \,
   M_a \cosh\theta\, L_a(\theta)
 +{\cal E}M_1^2\,L
\label{ecircl}
\en
where
${\cal E}M_1^2\,L$
  is the bulk contribution to the energy and
\eq
  \phi_{ab}(\theta)
= - {i \over 2 \pi}\frac{d}{d \te} \ln S_{ab}(\theta)\,.
\en
\subsection{ $(\alpha, \beta)$ boundary conditions:}
The (R-channel) thermodynamic Bethe ansatz equations are \cite{LMSS}
\bea
  \epsilon_a(\theta)
=
   2 M_a R \cosh\te -\ln \left( R_\alpha^{(a)}(i \fract{\pi}{2}-\te)
R_\beta^{(a)}(i \fract{\pi}{2} + \te) \right )
 - \sum_{b=1}^{N} \int_{\Rth} d \te'
   \phi_{ab}(\te - \te')\,
   L_b(\te')\,~~
\label{Ltba1}
\eea
where $a=1,\dots, N$\,; the ground state energy on an interval of
length $R$ is
then
\eq
   E_0^{\rm strip}(M,R) = -\sum_{a=1}^{N}
   \int_{\Rth} {d \te \over 4 \pi}
   M_a \cosh\theta\, L_a(\theta)\,
 +{\cal E}  M_1^2\, R
  + f_{\alpha} + f_{\beta}~,
\label{e0strip}
\en
where the constant ${\cal E}$ is the same as in (\ref{ecircl}),
$f_{\alpha}$ and $f_{\beta}$ are $R$-independent
contributions to the energy from the boundaries and
$\{ R_\alpha^{(a)}(\te), R_\beta^{(a)}(\te) \}$ are the
reflection amplitudes
corresponding to the two boundary conditions $\alpha$ and $\beta$.
Generalisations of these equations govern the
excited state energies $E_n^{\rm strip}(M,R)$
\cite{Us1}\footnote{Sometimes
such generalisations
are required even to describe the ground state correctly
\cite{Us1}, but
these cases will not concern us here.}, but
in the large $R$ limit we are interested in, they reduce to
simple (Bethe Ansatz) forms. Suppose that the $n^{\rm th}$ excited
state is made up of $m=\sum_{a=1}^Nm^{(a)}$ particles, $m^{(a)}$
being the number of particles of type $a$. Then
\eq
E_n^{\rm strip}(M,R)-E_0^{\rm strip}(M,R)= \sum_{a=1}^{N}
\sum_{i=1}^{\,m^{(a)}} M_a
\cosh \te_i^{(a)}  +O(e^{-RM})\,,
\label{e2}
\en
where sums on the RHS with $m^{(a)}=0$ are understood
to be omitted, and
the sets of numbers $\{ \te_i^{(a)} \}$  satisfy the
Bethe ansatz equations
\bea
2 \pi n_i^{(a)} &=& 2 M_a R \sinh \te_i^{(a)} -
i \ln \left (R_{\alpha}^{(a)}(\te_i^{(a)}) R_{\beta}^{(a)}(\te_i^{(a)})
\right)
\nn \\
&-&\sum_{b=1}^{N} \sum_{j \ne i} i
\ln \left (-S_{ab}(\te_i^{(a)}+ \te_j^{(b)}) \right) -
\sum_{b=1}^{N} \sum_{j \ne i} i \ln \left (-S_{ab}(\te_i^{(a)}-
\te_j^{(b)})  \right)\,.~~~~
\label{BAEQ1}
\eea
It is to be noted that the logarithmic branches in the Bethe ansatz
equations cause some difficulties in the numerics.
We impose the branch cut at  $-\pi$ so that
the function $-i \ln(RR) $ and  any  of the   terms $-i \ln(-S) $  in
(\ref{BAEQ1}) take values in the range  $(-\pi,\pi]$.  This
choice renders the Bethe ansatz (\ref{BAEQ1}) fully
anti-symmetric (a change of sign in any of the quantum numbers $n_i^{(a)}
\rightarrow -n_i^{(a)}$  corresponds to a change $\te_i^{(a)}
\rightarrow -\te_i^{(a)}$) and  one can consistently restrict
$\{ n_i^{(a)} \}$ to
strictly positive integers only.
\resection{Some exact results for the boundary Ising model}
\label{apps}
The boundary scattering matrix for the free Majorana fermion
is~\cite{GZ}
\eq
R_{k}(\theta)= i \tanh \left( {i \pi \over 4} -{\te \over 2}
\right) { k- i \sinh \te \over k+i \sinh \te}\,,
\label{Rkk}
\en
where $k=1-{ h^2 \over 2 M}\,$ and $h$ is  the boundary magnetic field.
Therefore, we have
\eq
\phi_k(\te)={ 1 \over \pi} \left({ 1 \over \cosh \te} -
{ 4 k \cosh \te \over \cosh(
2\te)+2 k^2 -1} \right)\,.
\label{phikk}
\en
In this appendix  we would like to report  exact expressions for
\eq
\ln g_{\bf free} \equiv \ln g_{k=1}(l) = -{ 1 \over 4} \int_{\Rth} d \te
 \left(   \delta(\te) + { 1 \over \pi \cosh \te} \right)
 \ln(1+ e^{-l \cosh \te})
\label{iex1}
\en
and
\eq
\ln g_{\bf fixed} \equiv \ln g_{k=-\infty}(l) = - { 1 \over 4}
\int_{\Rth} d \te
 \left( \delta(\te) -{ 1 \over\pi  \cosh \te}  \right)
 \ln(1+ e^{-l \cosh \te})\,.
\label{iex2}
\en
These are obtained from the following
identity,  which can be deduced  by studying the monodromies of the
integral~(\ref{iex}) along the lines sketched in~\cite{ex1}:
\eq
\int_{\Rth} d \te {1 \over \pi \cosh\te} \ln
\big( 1+e^{- x \cosh \te}\big )=
\ln(2) - {x \over \pi}(1 + \ln(\pi/x) -\gamma_{E})- 2  S(x)
\label{id1}
\en
where $\gamma_{E}=0.57721566 \dots$ is the Euler-Mascheroni constant and
\eq
S(x)= \sum_{n=1}^{\infty} \left(
\ln \left[ {x+ (2n-1) \pi -\sqrt{ (2n-1)^2 \pi^2 + x^2} \over x- (2n-1)
\pi +\sqrt{ (2 n-1)^2
\pi^2
+ x^2} }\right] + {x \over  (2 n-1) \pi} \right)\,.
\label{sss}
\en
The idea of the derivation is to determine the positions of the
singularities in (\ref{sss}) using the pinched-contour
argument of~\cite{ex1},  adding counterterms to make  the
infinite sum $S(x)$ convergent. The remaining parts of
(\ref{id1}) were then fixed by studying the $x \rightarrow 0$ limit.
(Alternatively, (\ref{id1})  can be proved using the
Bessel-function technique of \cite{KM,FR} and  an  identity
due to Schl{\"o}milch~\cite{GR}.) {}From (\ref{id1}), we have
\eq
4 \ln g_{\bf free/fixed}(l)=-
\ln(1+e^{-l}) \mp \left(\ln(2) - {l \over \pi}(1 +
\ln(\pi/l) -\gamma_{E})- 2  S(l) \right)\,.
\en
%
%
%%%%%%%%%%%%%%%%%%%%%%%%%%%%%%%%%
%

\end{document}